\documentclass[final,3p,times,twocolumn]{elsarticle}
\usepackage{amssymb}
\usepackage{amsmath}
\usepackage{xcolor}
\usepackage{fancyhdr}
\usepackage[T1]{fontenc}
\usepackage[utf8]{inputenc}   
\usepackage{textcomp}

\newcommand{\hmu}{\hat{\mu}}

\newcommand \xb {\bar{x}}

\begin{document}

\begin{frontmatter}

 \title{Studying the QCD phase diagram using pressure derivatives from lattice QCD}

 \author[1]{Sipaz Sharma}
 \ead{sipaz.sharma@tum.de}

\affiliation[1]{
  organization={Physik Department,~Technische~Universität~München},
  addressline={\\James\mbox{-}Franck\mbox{-}Straße~1},
  city={Garching~b.~München},
  postcode={D\mbox{-}85748},
  country={Germany}
}

\begin{abstract}
We summarize the application of derivatives of the QCD pressure, calculated within the framework of lattice QCD, in constructing observables that probe aspects of the QCD phase diagram at physical quark masses. We outline how the behavior of energy-like and magnetization-like observables at physical quark masses is influenced by the $(2+1)$-flavor chiral phase transition. We describe features of the chiral crossover at vanishing and non-vanishing chemical potentials and discuss deconfinement at zero chemical potential. We address the relevance of the convergence properties of the Taylor expansion of the QCD pressure in the search for the QCD critical endpoint.
\end{abstract}

%% Keywords
%\begin{keyword}



%\end{keyword}

\end{frontmatter}
\thispagestyle{fancy}
\fancyhf{} 
\fancyhead[R]{TUM-EFT 208/26} 
\renewcommand{\headrulewidth}{0pt} 



\section{Introduction}
\label{sec1}

\begin{figure}[h]
    \centering
    \includegraphics[width=\linewidth]{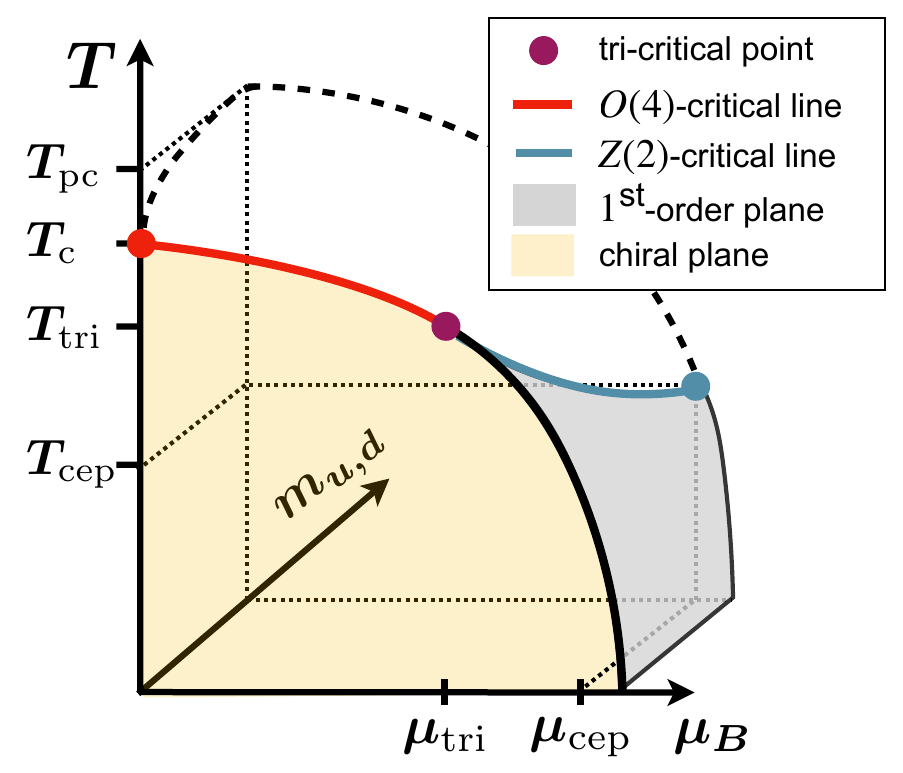}
    \caption{Sketch of the QCD phase diagram taken from Ref.~\cite{Karsch:2019mbv}. The backward plane is most relevant for this article as it represents the physical light quark mass.}
    \label{fig:phase1}
\end{figure}

Derivatives of the QCD pressure, $P$, with respect to temperature, quark masses, or the chemical potentials, $\mu_X$, of the conserved charges, $X$, are powerful tools for investigating the features of the QCD phase diagram \cite{Aarts:2023vsf} shown in Fig.~\ref{fig:phase1}. In the chiral limit of two massless quark flavors, the chiral symmetry group, $SU(2)_L\times SU(2)_R$, gets spontaneously broken down to the diagonal subgroup, $SU(2)_V$. At vanishing baryon chemical potential, $\mu_B$, this spontaneously broken chiral symmetry gets restored by a true second-order phase transition  at a critical temperature $T_c=132^{+3}_{-6}$MeV \cite{HotQCD:2019xnw}. However, the universality class of the two-flavor chiral phase transition depends upon the extent of the $U_A(1)$ breaking near the transition \cite{Pisarski:2024esv}, and the status of the latter near $T_c$ remains an open question. The signature of this genuine phase transition manifests at the physical value of the approximate degenerate up and down (or light) quark masses, $m_{u,d}\equiv m_l$, as a chiral crossover at a pseudocritical temperature, $T_{pc}=156.5\pm1.5$ MeV \cite{HotQCD:2018pds, Borsanyi:2020fev, Kotov:2021rah}. The chiral phase transition and the chiral crossover at vanishing $\mu_B$ have been studied on the lattice using derivatives of the $P$ involving at least one derivative w.r.t. the symmetry breaking field, $H=m_l/m_s$, such that the strange quark mass, $m_s$ is  always fixed to its physical value, whereas $m_l$ is varied from its physical value down to the chiral limit. The reason for the previous statement is that the order of the parameter of the chiral phase transition (chiral condensate) is related to the first derivative of $P$ w.r.t. $H$. 

As is well-known, at the finite values of $\mu_B$, a direct calculation using Monte Carlo is not possible because of the
infamous sign problem. Nevertheless, the critical, $T_c(\mu_B)$ and pseudocritical, $T_{pc}(\mu_B)$ phase boundaries emerging from $T_c$ and $T_{pc}$, respectively, can be constructed by Taylor expanding the order parameter and its various susceptibilities around $\mu_B=0$ as detailed in \cite{HotQCD:2019xnw, Ding:2024sux}. Additionally, the Taylor expansion coefficients of $P$ around vanishing chemical potentials, also known as generalized susceptibilities, have been used to produce the equation of state of the strongly interacting matter for the conditions met in the heavy-ion collision experiments \cite{Bollweg:2022fqq}. The convergence properties of the Taylor series of pressure can be used to put bounds on the location of the most searched QCD critical endpoint expected to belong to the $3d$-$Z_2$ Ising universality class \cite{Hatta:2002sj, Stephanov:2006dn}.  Although there is no sign problem at the imaginary chemical potential, the fit function choice for
analytic continuation is not unique, and the exploration is only possible up to a
certain value as the Roberge-Weiss phase transition occurs at $\mu^2=-(\pi/3)^2$ \cite{DElia:2002tig, deForcrand:2002hgr, Borsanyi:2012cr, Guenther:2017hnx}. For more methods on attempting to curb the sign problem, such as reweighting techniques, density of state methods,
using the canonical ensemble, formulations with dual variables,
Lefschetz thimbles, complex Langevin, refer to the review \cite{Guenther:2020jwe}, and references therein. 

At low temperatures
and low baryon densities, strongly interacting matter is confined, and the relevant degrees of freedom are hadrons,
which are composite states of the fundamental particles—quarks and gluons. At low baryon densities, the chiral
crossover and the creation of QGP from hadrons are commonly expected to occur at the same temperature. However, the deconfinement aspect is not well-understood in lattice QCD studies involving light quarks. The generalized susceptibilities involving the heavier charm quark have been proven to be a good probe of the deconfinement transition \cite{Bazavov:2023xzm, Kaczmarek:2025dqt}.

In Section 2, we will discuss the importance of the universal scaling framework in order to understand the behaviour of derivatives of the QCD pressure at physical quark masses. In Section 3, we will review our current understanding of the chiral crossover line and its relation to the experimental results based on the chemical freeze-out. We will also discuss the onset of deconfinement at $T_{pc}(\mu_B=0)$ from the lens of the charm sector. In Section 4, we will review the convergence properties of the Taylor series of the pressure and its relation to the QCD critical point. We will give concluding remarks in the last section.
 
\section{Universal scaling framework applied to the Taylor expansion coefficients of the pressure}
\label{sec2}
\begin{figure*}[h]
      \includegraphics[width=0.5\textwidth]{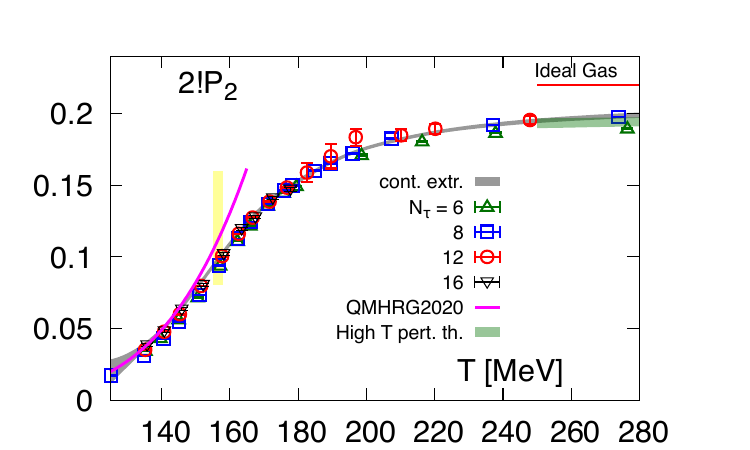} 
      \includegraphics[width=0.5\textwidth]{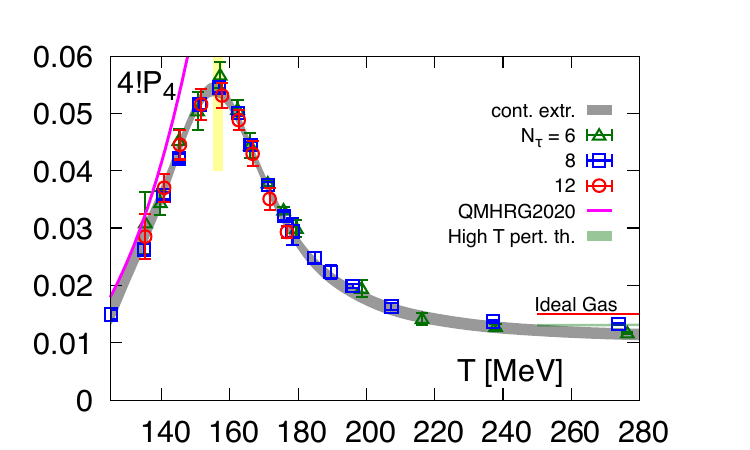} \\
      \includegraphics[width=0.5\textwidth]{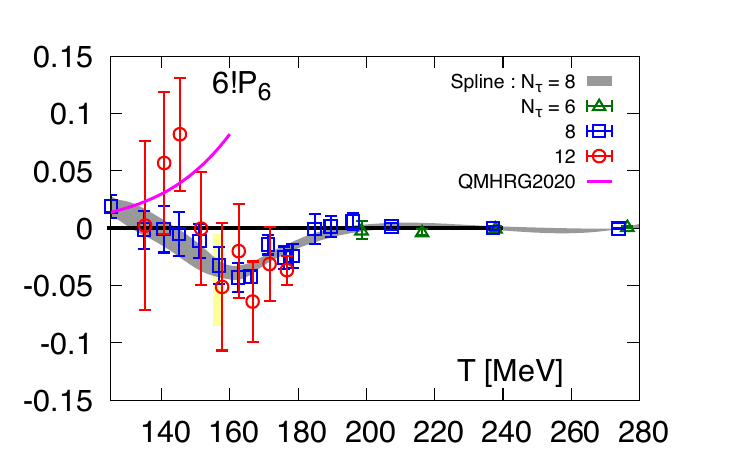} 
        \includegraphics[width=0.5\textwidth]{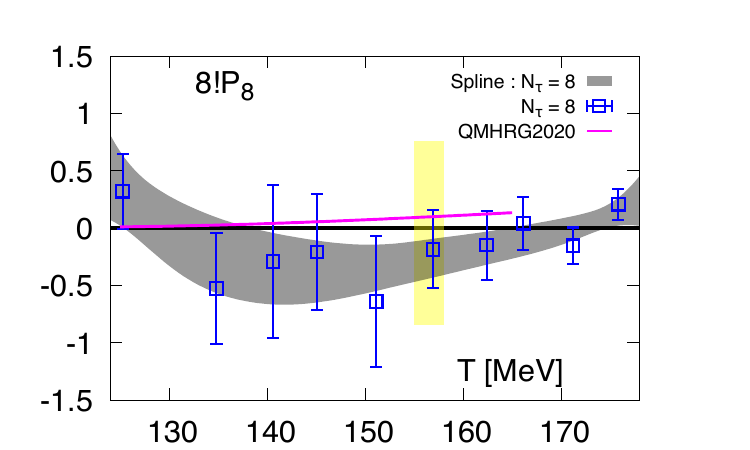}
        \caption{Shown are the Taylor expansion coefficients of the QCD pressure defined in Eq.\eqref{eq:6} for the strangeness-neutral and isospin-symmetric matter. The figures are taken from Ref.~\cite{Bollweg:2022fqq}.}
        \label{fig:P2n}
 \end{figure*}
In the vicinity of a second-order critical point or line or a surface, the free energy density, 
$f$
%= - (T/V)\ln Z$, 
can be split into a dominant
singular contribution and  sub-leading\footnote{The sub-leading terms comprise regular terms and corrections to scaling.} corrections that are of relevance in some range of $H\ne 0$ \cite{Ding:2024sux},
  
 \begin{equation}
 f = {h_0 h^{1+1/\delta} f_f(z)} + \text{ sub-leading} \;  .  
 \end{equation}

In the equation above, the infinite volume scaling function, $f_f(z)$, only depends on a single scaling variable, $z=z_0 z_b$, with
 $z_b=\bar{t}/H^{1/\beta\delta}$ and $ z_0 = h_0^{1/\beta\delta}/t_0$. Here, $t_0$, $h_0$, $T_c$ are the non-universal parameters, whereas $\beta$ and $\delta$ are the critical exponents of the universality class of the relevant second-order phase transition. The scaling variable depends upon the energy-like scaling fields, $t$, that do not break the underlying symmetry, and the magnetization-like scaling fields, $h$, that break the underlying symmetry. For example, in the context of chiral symmetry, temperature, $T$, and chemical potentials of conserved charges are energy-like couplings, whereas quark mass breaks the chiral symmetry explicitly and is classified as a magnetization-like coupling. These fields couple to energy-like and magnetization-like observables in the QCD Lagrangian.
  \begin{eqnarray}
  t&=&\frac{\bar{t}}{t_0}= \frac{1}{t_0} \left( \frac{T-T_c}{T_c} + \kappa_2^B \hmu_B^2 
  + ...\right), \nonumber \\
h&=&\frac{H}{h_0} = \frac{1}{h_0} \frac{m_\ell}{m_s}. 
  \label{eq:coupling-h}
  \end{eqnarray}
The curvature coefficient, $\kappa_2^B$, is defined at  $T_c$, and is also a non-universal parameter. We have adapted the standard convention to use dimensionless $\hmu_X=\mu/T$.

In a grand canonical ensemble, 
\begin{equation}
    f=-P=-\dfrac{T}{V}\ln\mathcal{Z} \; .
\end{equation}
The pressure in turn can be expanded around $\vec{\mu}=~(\hmu_B,\hmu_Q,\hmu_S,\hmu_C)=0$, where $B, Q, S$ and $C$ denote the net baryon number, electric charge, strangeness, and charm quantum numbers:

\begin{eqnarray}
\hat{P}\equiv P/T^4 &=& \frac{1}{VT^3}\ln\mathcal{Z}(T,V,\hmu_B,\hmu_Q,\hmu_S, \hmu_C) \nonumber\\
&=& \sum_{i,j,k,l=0}^\infty
\frac{\chi_{ijkl}^{BQSC}}{i!j!k!l!} \hmu_B^i \hmu_Q^j \hmu_S^k \hmu_C^l \; ,
\label{eq:pressure}
\end{eqnarray}
where the expansion coefficients, $\chi_{ijkl}^{BQSC}$, are the so-called generalized susceptibilities\footnote{In the following, whenever a subscript is zero, it will not be written explicitly. The same applies to the corresponding superscript.} (or charge fluctuations and correlations) given by,
\begin{eqnarray}
\chi_{ijkl}^{BQSC}\equiv \chi_{ijkl}^{BQSC}(T) &=& \left. 
\frac{\partial \hat{P}}{\partial\hmu_B^i \partial\hmu_Q^j \partial\hmu_S^k \partial \hmu_C^l}\right|_{\vec{\mu}=0}.
\label{eq:chi}
\end{eqnarray}

In the heavy-ion collisions of gold or lead,  the net strangeness density, $n_S$, vanishes, whereas the ratio of net electric-charge density ($n_Q$) to net baryon number density ($n_B$) is close to 0.4. However, it was shown in \cite{Bazavov:2017dus} that the thermodynamic quantities do not vary much with $n_Q/n_B$, so the isospin-symmetric case $n_Q/n_B=0.5$ is also a good approximation. Given these constraints, $\hmu_Q$ and $\hmu_S$ can be expressed as Taylor series in $\hmu_B$. We note that the charm quarks are formed in the initial-hard thermal collisions, so expansion in $\hmu_C$ is not relevant for comparison with the available experimental results. For the strangeness neutral ($n_S=0$) and isospin symmetric matter ($n_Q/n_B=0.5$), one obtains,

\begin{eqnarray}
    \hat{P} &=& \sum_{k=0}^\infty {P_{2k}}\hat{\mu}_B^{2k}, \nonumber \\
\text{with }\;\;\;\; P_{2k} &=& \frac{\Tilde{\chi}^B_{2k}(T)}{2k!} .
\label{eq:6}
 \end{eqnarray}
For the definition of $\Tilde{\chi}^B_{2k}(T)$, see Ref.~\cite{Bollweg:2022fqq}, and Appendix A of Ref.~\cite{Bazavov:2020bjn}\footnote{We refer to Refs.~\cite{Bollweg:2022fqq,Bazavov:2020bjn} for the explicit expressions of the constrained Taylor expansion coefficients, as they are rather lengthy and the discussion in this work does not rely on their detailed form.}. Obviously, $H$ attains its physical value in the experiments, and the results reviewed in the following will be based on $(2 + 1)$-flavor ensembles \cite{Bollweg:2021vqf} of the HotQCD collaboration
with two degenerate light and a heavier strange quark tuned to their physical mass
values while keeping $m_\ell/m_s=1/27$. These data sets have been generated
using the HISQ (Highly Improved Staggered Quark Action) action with tree-level coefficients
and a tree-level improved gauge action. For details on the ensembles and calculation, refer to \cite{Bollweg:2022fqq}.

In the scaling regime, given the form of energy-like scaling field in Eq.~\eqref{eq:coupling-h}, one derivative w.r.t. $T$ of a function $g$ should be proportional to its second-order derivative w.r.t. $\hmu_B$. 
\begin{equation}
\left. \frac{\partial^2g}{\partial \hmu_B^2}\right|_T=\left. 2\kappa_2^B T_c\frac{\partial g}{\partial T}\right|_T
\label{eq:derivative}
\end{equation}
In Fig.~\ref{fig:P2n}, the four Taylor expansion coefficients of the pressure up to $\mathcal{O}(\hmu_B^8)$, as defined in Eq.~\eqref{eq:6}, follow the scaling expectation of Eq.~\eqref{eq:derivative}. This implies that the behaviour of the energy-like observables shown in Fig.~\ref{fig:P2n} can be understood using the universal scaling arguments, and at the same time, $\kappa_2^B$ can be extracted from them. For example, $P_4$ behaves like the specific heat that should develop a pronounced cusp in the chiral limit, but no divergence as the critical exponent $\alpha$ of a $3d$-$O(N)$\footnote{We note that even though there are indications that the 2-flavor chiral phase transition belongs to $3d$-$O(4)$ universality class, this needs further clarification.} universality class is negative with magnitude less than 1, and should receive a larger regular contribution at $T_c$.
\section{Chiral crossover and deconfinement}
\label{sec3}
\begin{figure}[h]
\includegraphics[width=0.5\textwidth]{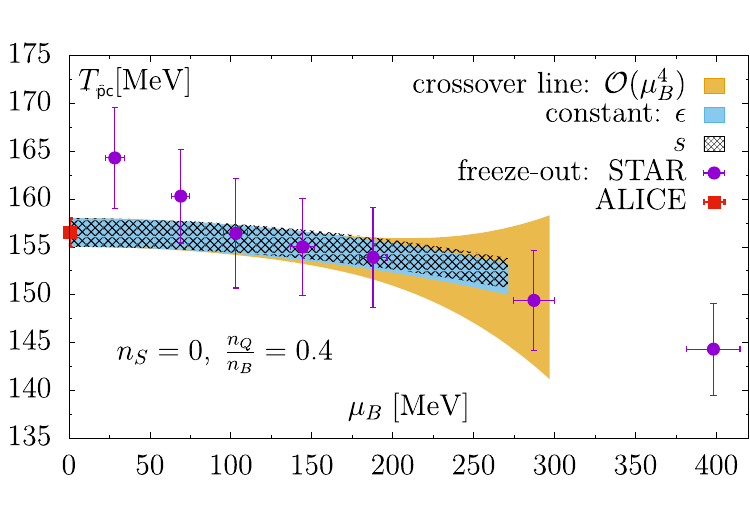} 
   
    \caption{Shown is the comparison of the pseudocritical phase boundary obtained from the lattice QCD calculations and the freeze-out parameters obtained by the experiments. The figure is taken from Ref.~\cite{HotQCD:2018pds}.}
     \label{fig:Tpc}
\end{figure}
In Ref.~\cite{HotQCD:2018pds}, various magnetization-like observables were used to calculate the continuum pseudocritical temperature corresponding to the chiral crossover at physical quark masses and vanishing $\hmu_B$. These observables were related to the following types of pressure derivatives:
\begin{eqnarray}
    \chi^{M_\ell}_{t(T)} &\sim& -\frac{\partial^2 P}{\partial T \partial m_\ell}, \nonumber \\
    \chi^{M_\ell}_{t(\ell,\ell)}
&\sim& - \frac{\partial^3 P}{\partial \hmu_\ell^2\partial m_\ell}, \nonumber\\
\chi_{m} &\sim& \frac{\partial^2 P}{\partial m_\ell^2}, 
\label{eq:mag}
\end{eqnarray}
$\hmu_\ell$ is a flavor chemical potential. For the isospin symmetric case, $\hmu_\ell=\hmu_B/3$. $\chi^{M_\ell}_{t(T)}$ and $\chi^{M_\ell}_{t(\ell,\ell)}$ are related to the derivative of the scaling function of the order parameter, $-f_G'(z)$, that peaks at a characteristic value $z_t$, whereas $\chi_{m}$ is related to the scaling function of the susceptibility of the order parameter, $f_\chi(z)$, that peaks at $z_m$ \cite{Karsch:2023pga}. The latter is a pure magnetization-like observable, while the remaining two are mixed observables. Ignoring the sub-leading corrections, different types of pseudocritical temperatures ($T_{pc,t/m}$) obtained from different types of observables are related to the critical temperature as follows: 
\begin{equation}
\small T_{pc,t/m}(\hmu_B, H)= 
T_c \left( 1 + (\kappa_2^B\hmu_B^2+...) + \frac{z_{t/m}}{z_0} H^{1/\beta\delta}\right)
\end{equation}
As can be seen from the previous equation, the pseudocritical temperatures are not unique and depend upon the observable used to calculate them, and they should all converge to a unique value in the chiral limit. In addition to the continuum $T_{pc}(\hmu_B=0)$, obtained from a weighted average of various pseudocritical temperatures, Fig.~\ref{fig:Tpc} also shows the pseudocritical phase boundary as a function of $\hmu_B$ for the conditions met in the heavy-ion collision experiments. The lattice results of the phase boundary in Fig.~\ref{fig:Tpc} were constructed by extracting the curvature coefficient, $\kappa_2^B=0.012(4)$, using magnetization-like and mixed susceptibilities that make use of the pressure derivatives listed in Eq.\eqref{eq:mag}. Ref.~\cite{Bollweg:2022fqq} found that the curvature coefficient extracted from the energy-like observables discussed in Section \ref{sec2} is in agreement with its value based on magnetization-like observables. We note that the $\mathcal{O}(\hmu_B^4)$ curvature coefficient has been found to be insignificant in lattice  QCD investigations \cite{Bhattacharya:2014ara, HotQCD:2018pds}.
\begin{figure}[h]
 \includegraphics[width=0.5\textwidth]{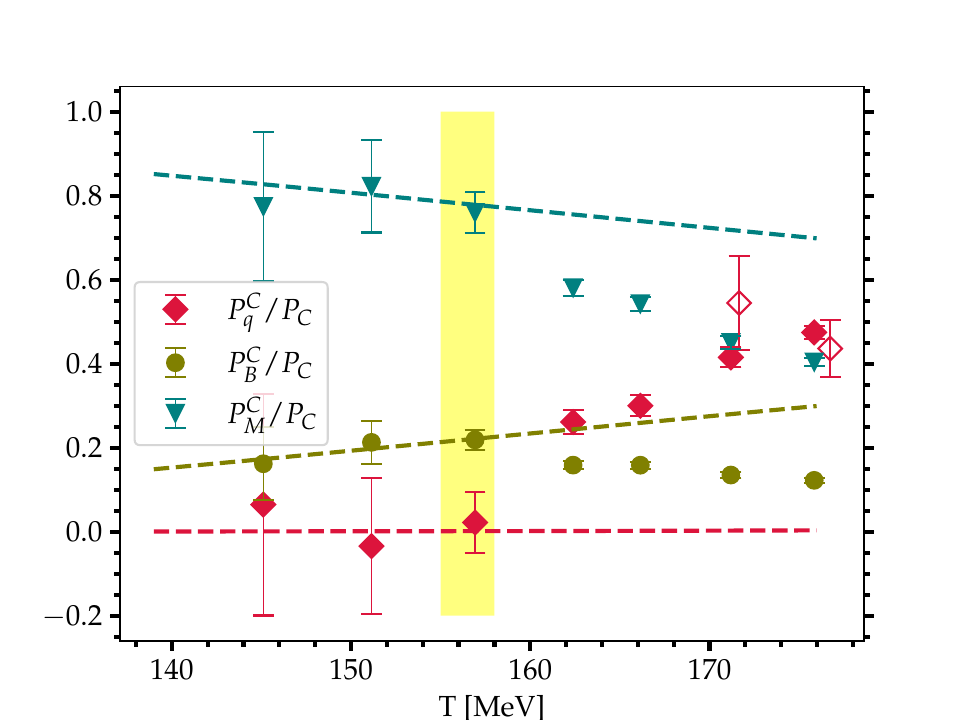} 
   
    \caption{Shown are the relative contributions of charm quark-like excitations $P_q^{C}$, charmed baryons, $P_B^C$, and charmed mesons, $P_M^C$ to the partial charm pressure, $P_C$. The dashed lines of the same color represent the corresponding HRG predictions. The yellow band represents $T_{pc}$ with its uncertainty. The figure is taken from Ref.~\cite{Bazavov:2023xzm}.}
     \label{fig:Pqc}
\end{figure}

Charge fluctuations and correlations measured in heavy-ion collisions are expected to originate at the chemical freeze-out temperature $T_f$, the stage at which inelastic interactions cease and particle abundances become fixed. The corresponding freeze-out parameters characterize the thermal conditions at this point but cannot be accessed directly in experiments. Instead, freeze-out parameters are extracted through model-dependent analyses that compare measured particle yields with the hadronization models \cite{Andronic:2017pug}. As can be seen from Fig.~\ref{fig:Tpc}, the resulting beam-energy dependence of $T_f(\sqrt{s_{NN}})$ is found to be consistent with the pseudo-critical temperature $T_{pc}(\hat{\mu}_B)$ obtained from the lattice QCD calculations.

This agreement supports the use of hadron resonance gas (HRG) models as an effective description of strongly interacting matter near freeze-out. In the low temperature phase where HRG is valid, the fact that the Boltzmann approximation is valid for relatively heavier open-charm hadrons (as well as the charm quark), makes it easier to construct partial pressure contributions of charmed hadron pressure to QCD pressure using generalized susceptibilities involving charm \eqref{eq:pressure}.  It was shown in Ref.~\cite{Bazavov:2023xzm} that the lattice results on generalized charm susceptibilities agree with their HRG counterparts in the low temperature phase, and deviate from HRG at the chiral crossover. The breakdown of HRG at $T_{pc}$ signals the onset of the melting of the charmed hadrons and appearance of charm quark-like degrees of freedom inside QGP. The non-interacting gas of charmed hadrons was extended to include charm-quark-like excitations inside QGP, and was found to be valid in Ref.~\cite{Bazavov:2023xzm}. In this quasi-particle model, one can construct proxies for the partial pressure of charm quark, $P_q^{C}$, charmed baryons, $P_B^C$, and charmed mesons, $P_M^C$ as follows:
\begin{eqnarray}
		P_{q}^{C}& =&9(\chi^{BC}_{13}-\chi^{BC}_{22})/2\\
		 P_{B}^{C}&=&(3\chi^{BC}_{22}-\chi^{BC}_{13})/2\\
		 P_{M}^{C}& =&\chi^{C}_{4}+3\chi^{BC}_{22}-4\chi^{BC}_{13}
\end{eqnarray}
As can be seen in Fig.~\ref{fig:Pqc}, relative contributions of both $P_{B}^{C}$ and $P_M^{C}$ to the partial charm pressure $P_C\equiv\chi^C_4$ agree well with the HRG calculation, that takes into account experimentally observed charmed hadrons and the ones predicted by quark model, up to $T_{pc}$. Above $T_{pc}$, the share of charmed hadrons to $P_C$ starts dropping, and the relative contribution of $P_q^C$ to $P_C$ starts rising. $P_q^C$ becomes the dominant contributor above 175 MeV \cite{Sharma:2024edf}.These results indicate that at $\hmu_B=0$, deconfinement coincides with the chiral crossover. Moreover, based on the comparison of the extended HRG model explained above to the continuum lattice results \cite{Sharma:2025gtg}, Ref.~\cite{Kaczmarek:2025dqt} predicted that the experimentally unobserved charmed baryons give approximately 49\% contribution to the partial charmed baryon pressure at the freeze-out. In contrast, the charmed meson sector is experimentally well-known, and approximately 12\% contribution to $P_M^C$ comes from the experimentally unknown states \cite{Sharma:2025zhe}. These results are also corroborated by the analysis of the open-charm hadron yield observed in experiments \cite{Braun-Munzinger:2024ybd}.

 \section{Taylor expansions of thermodynamic quantities and the QCD critical endpoint}
 \begin{figure*}[h]
      \includegraphics[width=0.50\textwidth]{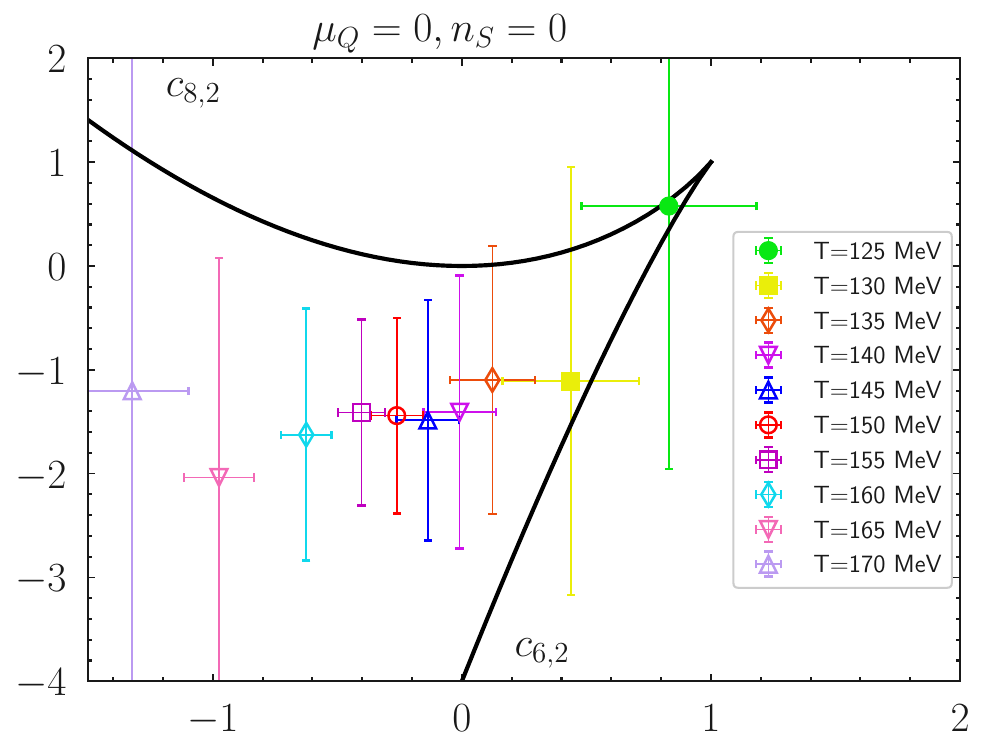} 
      \includegraphics[width=0.55\textwidth]
     {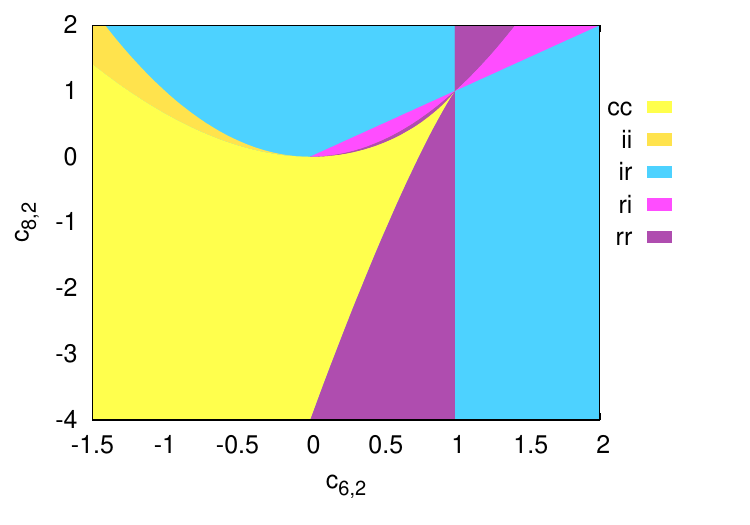} 
        \caption{[{\it Right}] The figure maps the pole structure of the [4,4] Padé approximant expressed in terms of two Taylor expansion parameters in Eq.~\eqref{eq:pade}. [{\it Left}] The figure shows the location of  lattice results at various temperatures in the pole map. The figures are taken from Ref.~\cite{Bollweg:2022rps}.}
        \label{fig:pade}
 \end{figure*}
Radius of convergence of a Taylor series around $\hmu_B=0$ is related to a phase transition if the breakdown of the series expansion is caused by a singularity that is closest to the origin and lies on the real $\hmu_B$ axis \cite{Karsch:2010hm}. In the context of QCD, such a real singularity can be related to the QCD critical point. However, in practical lattice QCD calculations, only a few Taylor expansion coefficients are available, and thus one can calculate an estimator of the radius of convergence and some information about the analytic structure of the Taylor expansion using the limited number of coefficients. In particular, using Taylor expansion coefficients of the QCD pressure defined in Eq.~\eqref{eq:6}, the following estimator of the radius of convergence can be obtained,
\begin{equation}
r_{2k} = \displaystyle\lim_{k \to \infty} \left| \frac{P_{2k}}{P_{2k+2}} \right|^{1/2} .
\label{eq:radius}
\end{equation}
Although the limit, $k\rightarrow\infty$, cannot be realized in practice, successive ratios of the available coefficients provide approximate estimates of the radius of convergence and thus yield indirect information about the analytic structure of the thermodynamic potential in the complex $\hmu_B$ plane. 

The convergence of the Taylor series of the pressure can be improved by employing resummation techniques. In particular, the Padé resummation technique utilizes the same derivative information that enters the Taylor expansion coefficients, but reorganizes it into a rational function. This approach leads to an improved convergence behavior, particularly in the vicinity of singularities. In many cases, reliable results can be obtained with a smaller number of input coefficients compared to a truncated Taylor series \cite{Clarke:2024ugt}. Since poles and other singular structures are more naturally represented by rational functions than by finite-order polynomials, Padé approximants are especially well suited for capturing non-analytic features of thermodynamic observables. As a consequence, they provide better characterization of the poles and can extend the applicability of results beyond the estimated radius of convergence of the original Taylor expansion. 

In Ref.~\cite{Bollweg:2022rps}, the $\mathcal{O}(\hmu_B^8)$ Taylor series of the $\hmu_B$-dependent part of the pressure, $\Delta P (T,\hmu_B)$, is rearranged such that the convergence properties become dependent only on two expansion coefficients,
\begin{eqnarray}
\frac{(\Delta P(T,\mu_B)/T^4) P_4}{P_2^2} &=& 
\sum_{k=1}^{\infty} c_{2k,2} \xb^{2k}\; 
\\
&=& \xb^2+\xb^4+ c_{6,2} \xb^6
+ c_{8,2} \xb^8 + ...\;, \nonumber
\label{eq:rearranged}
\end{eqnarray}	
\begin{equation}
    \text{where }\;\; \xb=\sqrt{\frac{P_{4}}{P_2}}\ \hmu_B , \; \;
 c_{2k,2}=\frac{P_{2k}}{P_2} \left(\frac{P_2}{P_{4}}\right)^{k-1}\; .   \nonumber
\end{equation}
The following [4,4] Padé approximant of the Taylor series, 
\begin{equation}
 P[4,4] = \frac{(1-c_{6,2}) \xb^2 +
\left(1 - 2 c_{6,2} +c_{8,2} \right) \xb^4}{(1-c_{6,2})+
(c_{8,2}-c_{6,2}) \xb^2 + (c_{6,2}^2 - c_{8,2}) \xb^4} , 
\label{eq:pade}
\end{equation}
has four poles which come in two pairs:
\begin{equation}
z^\pm =\frac{c_{8,2} - c_{6,2} \pm \sqrt{
(c_{8,2}-c_{8,2}^+) (c_{8,2}-c_{8,2}^-)}}{2 (c_{8,2} -c_{6,2}^2)} \;\; , \nonumber
\end{equation}
\begin{equation}
\text{with}\;\;c_{8,2}^\pm = -2 + 3 c_{6,2} \pm 2(1-c_{6,2})^{3/2}
\;\; . \nonumber
\end{equation}
Fig.~\ref{fig:pade} [{\it Right}] maps the types of the poles in the parameter space of $c_{8,2}$ and $c_{6,2}$. In the light yellow region, denoted as cc, all four poles are complex. This same region is shown as bounded by the solid black lines in Fig.~\ref{fig:pade} [{\it Left}]. As can be seen, given the errors, except for the highest (170 MeV) and the lowest (125 MeV) temperatures, the nearest singularity to the origin is not purely real, and thus is not related to a phase transition at these temperatures. Fig.~\ref{fig:pade} [{\it Right}] also shows regions in the parameter space for the occurrence of two pairs of imaginary poles: ii; two pairs of real poles: rr; and one pair of imaginary and one pair of real poles: ir and ri, with the first letter indicating the type of pole pair closest to the origin. This means that if the error on the $T=125$ MeV point could be reduced, and at the same time if the 8th order coefficient becomes positive, the green point could move out of the region of four complex poles to a region where the closest singularity to the origin is real. It is worth pointing out that the Fig.~\ref{fig:pade} [{\it Left}] is older than the bottom-right panel of Fig.~\ref{fig:P2n}. Notice that within errors, the eighth order coefficient is positive in Fig.~\ref{fig:P2n}. Ultimately, these results can be used to put bounds on the location of the QCD critical endpoint. The sketch of the QCD phase diagram in Fig.~\ref{fig:phase1} depicts that $T_c(\hmu_B=0)=132^{+3}_{-6}$~MeV (the red dot) puts an upper bound on the QCD critical endpoint (the blue dot). The information extracted from a [4,4] Pade approximant is therefore consistent with this. For details on various estimates of the radius of convergence as a function of temperature, see Ref~\cite{Bollweg:2022rps}.
\section{Conclusions}
We have summarized how derivatives of the QCD pressure provide a common framework for studying multiple aspects of the QCD phase diagram. At physical quark masses,  both magnetization-like and energy-like observables constructed from pressure derivatives have imprints of the universal scaling properties related to the $(2+1)$-flavor chiral phase transition. At physical quark masses, this allows for a controlled determination of the pseudocritical temperature and the curvature of the chiral crossover line. For the determination of the curvature of the critical line in the chiral limit, see Ref.~\cite{Ding:2024sux}. 

The resulting crossover line is consistent with phenomenological extractions of chemical freeze-out parameters from heavy-ion collision experiments. In addition, generalized charm susceptibilities offer complementary information on deconfinement. Their behavior indicates that the melting of charmed hadrons and the appearance of charm quark-like degrees of freedom start at the chiral crossover temperature, suggesting that at vanishing baryon chemical potential, restoration of the spontaneously broken chiral symmetry group and deconfinement take place at the same temperature.

Finally, the convergence properties of the Taylor expansion of the pressure, supplemented by Padé resummation analyses, provide insight into the analytic structure of thermodynamic observables in the complex $\hat{\mu}_B$ plane. Current results do not indicate the presence of a nearby real singularity for temperatures around and above the chiral crossover, but they allow constraints to be placed on the possible location of a critical endpoint. 

We emphasize that all results discussed in this work are taken from previously published studies. The novelty of this manuscript lies in providing a unified perspective that connects these results through derivatives of the QCD pressure, rather than in presenting new simulation data.
\section*{Acknowledgements}
 We acknowledge support from the DFG cluster of excellence ORIGINS, funded by the Deutsche Forschungsgemeinschaft under Germany’s Excellence Strategy-EXC-2094-390783311. 
\bibliographystyle{elsarticle-num} 
\bibliography{refs}

\end{document}